# Detection of Individual Gas Molecules Adsorbed on Graphene

F. Schedin[1], A.K. Geim[1], S.V. Morozov[2], E.W. Hill[1], P. Blake[1], M.I. Katsnelson[3] & K.S. Novoselov[1]

[1]Manchester Centre for Mesoscience and Nanotechnology,
University of Manchester, M13 9PL, Manchester, UK
[2]Institute for Microelectronics Technology, 142432 Chernogolovka, Russia
[3]Institute for Molecules and Materials, University of Nijmegen, 6525 ED Nijmegen, Netherlands

**The ultimate aspiration of any detection method is to achieve such a level of sensitivity that individual quanta of a measured value can be resolved. In the case of chemical sensors, the quantum is one atom or molecule. Such resolution has so far been beyond the reach of any detection technique, including solid-state gas sensors hailed for their exceptional sensitivity [1-4]. The fundamental reason limiting the resolution of such sensors is fluctuations due to thermal motion of charges and defects [5] which lead to intrinsic noise exceeding the sought-after signal from individual molecules, usually by many orders of magnitude. Here we show that micrometre-size sensors made from graphene are capable of detecting individual events when a gas molecule attaches to or detaches from graphene's surface. The adsorbed molecules change the local carrier concentration in graphene one by one electron, which leads to step-like changes in resistance. The achieved sensitivity is due to the fact that graphene is an exceptionally low-noise material electronically, which makes it a promising candidate not only for chemical detectors but also for other applications where local probes sensitive to external charge, magnetic field or mechanical strain are required.**

Solid-state gas sensors are renowned for their high sensitivity, which – in combination with low production costs and miniature sizes – have made them ubiquitous and widely used in many applications [1,2]. Recently, a new generation of gas sensors have been demonstrated using carbon nanotubes and semiconductor nanowires (see, for example, refs [3,4]). The high acclaim received by the latter materials is, to a large extent, due to their exceptional sensitivity allowing detection of toxic gases in concentrations as small as 1 part per billion (ppb). This and even higher levels of sensitivity are sought for industrial, environmental and military monitoring.

The operational principle of graphene devices described below is based on changes in their electrical conductivity $\sigma$ due to gas molecules adsorbed on graphene's surface and acting as donors or acceptors, similar to other solid-state sensors [1-4]. However, the following characteristics of graphene make it possible to increase the sensitivity to its ultimate limit and detect individual dopants. First, graphene is a strictly two-dimensional material and, as such, has its whole volume exposed to surface adsorbates, which maximizes their effect. Second, graphene is highly conductive, exhibiting metallic conductivity and, hence, low Johnson noise even in the limit of no charge carriers [6-9], where a few extra electrons can cause notable relative changes in carrier concentration $n$. Third, graphene has few crystal defects [6-10], which ensures a low level of excess ($1/f$) noise caused by their thermal switching [5]. Fourth, graphene allows four-probe measurements on a single-crystal device with electrical contacts that are Ohmic and have low resistance. All these features contribute to make a unique combination that maximizes the signal-to-noise ratio to a level sufficient for detecting changes in a local concentration by less than one electron charge $e$ at room temperature.

The studied graphene devices were prepared by micromechanical cleavage of graphite at the surface of oxidized Si wafers [7]. This allowed us to obtain graphene monocrystals of typically ten microns in size. By using electron-beam lithography, we made electrical (Au/Ti) contacts to graphene and then defined multiterminal Hall bars by etching graphene in an oxygen plasma. The microfabricated devices (upper inset in Fig. 1a) were placed in a variable temperature insert inside a superconducting magnet and characterised by using field-effect measurements at temperatures $T$ from 4 to 400K and in magnetic fields $B$ up to 12T. This allowed us to find mobility $\mu$ of charge carriers (typically, $\approx 5,000$ cm$^2$/Vs) and distinguish between single-, bi- and few-layer devices, in addition to complementary measurements of their thickness carried out by optical and atomic force microscopy [6-9]. The lower inset of Fig. 1a shows an example of the field-effect behaviour exhibited by our devices at room $T$. One can see from this plot that longitudinal ($\rho_{xx}$) and Hall ($\rho_{xy}$) resistivities are symmetric and anti-symmetric functions of gate voltage $V_g$, respectively. $\rho_{xx}$ exhibits a peak at zero $V_g$ whereas $\rho_{xy}$ simultaneously passes through zero, which shows that the transition from electron to hole transport occurs at zero $V_g$ indicating that graphene is in its pristine, undoped state [6].

To assess the effect of gaseous chemicals on graphene devices, the insert was evacuated and then connected to a relatively large (5 litre) glass volume containing a selected chemical strongly diluted in pure helium or nitrogen at atmospheric pressure. Figure 1b shows the response of zero-field resistivity $\rho = \rho_{xx}(B=0) = 1/\sigma$ to $NO_2$, $NH_3$, $H_2O$ and CO in concentrations $C$ of 1 part per million (ppm). One can see large, easily detectable changes that occurred within 1 min and, for the case of $NO_2$, practically immediately after letting the chemicals in. The initial rapid response was followed by a region of saturation, in which the resistivity changed relatively slowly. We attribute this region to redistribution of adsorbed gas molecules between different surfaces in the insert. After a near-equilibrium state was reached, we evacuated the container again, which led only to small and slow changes in $\rho$ (region III in Fig. 1b), indicating that adsorbed molecules were strongly attached to the graphene devices at room $T$. Nevertheless, we found that the initial undoped state could be recovered by annealing at 150ºC in vacuum (region IV). Repetitive exposure-annealing cycles showed no "poisoning" effects of these chemicals (that is, the devices could be annealed back to their initial state). A short-time UV illumination offered an alternative to thermal annealing.

To gain further information about the observed chemical response, we simultaneously measured changes in $\rho_{xx}$ and $\rho_{xy}$ caused by gas exposure, which allowed us to find directly a) concentrations $\Delta n$ of chemically induced charge carriers, b) their sign and c) mobilities. The Hall measurements revealed that $NO_2$, $H_2O$ and iodine acted as acceptors whereas $NH_3$, CO and ethanol were donors. We also found that, under the same exposure conditions, $\Delta n$ depended linearly on concentration $C$ of an examined chemical (see Fig. 1a). In order to achieve the linear conductance response we electrically biased our devices (by more than ±10V) to higher-concentration regions, away from the neutrality point (NP), so that both $\sigma = ne\mu$ and Hall conductivity $\sigma_{xy} = 1/\rho_{xy} = ne/B$ were proportional to $n$ (see lower inset of Fig. 1a) [6-9]. The linear response as a function of $C$ should greatly simplify the use of graphene-based sensors in practical terms.

Chemical doping also induced impurities in graphene in concentrations $N_i = \Delta n$. However, despite these additional scatterers, we found no notable changes in $\mu$ even for $N_i$ exceeding $10^{12} cm^{-2}$. Figure 2 illustrates this unexpected observation by showing the electric field effect in a device repeatedly doped with $NO_2$. One can see the V-shaped $\sigma(V_g)$-curves characteristic for graphene [6-9]. Their slopes away from NP provide a measure of impurity scattering (so-called field-effect mobility $\mu = \Delta\sigma/\Delta ne = \Delta\sigma/e\alpha\Delta V_g$). The chemical doping only shifted the curves as a whole, without any significant changes in their shape, except for the fact that the curves became broader around NP (the latter effect is discussed in Supplementary Information). The parallel shift unambiguously proves that the chemical doping did not affect scattering rates. Complementary measurements in magnetic field showed that the Hall-effect mobility $\mu = \rho_{xy}/\rho_{xx}B$ was also unaffected by the doping and exhibited values very close to those determined from the electric field effect. Further analysis yields that chemically-induced ionized impurities in graphene in concentrations $>10^{12}$ cm$^{-2}$ (that is, less than 10 nm apart) should not be a limiting factor for $\mu$ until it reaches values of the order of $10^5$ cm$^2$/Vs, which translates into the mean free path as large as $\approx 1\mu$m (see Supplementary Information). This is in striking contrast with conventional 2D systems, in which so high densities of charged impurities are detrimental for ballistic transport, and also disagrees by a factor of >10 with recent theoretical estimates for the case of graphene [11-13]. Our observations clearly raise doubts about charged impurities being the scatterers that currently limit $\mu$ in graphene [11-13]. In Supplementary Information, we show that a few-nm-thick layer of absorbed water provides sufficient dielectric screening to explain the suppressed scattering on charged impurities. We also suggest there that microscopic corrugations of a graphene sheet [14,15] could be dominant scatterers.

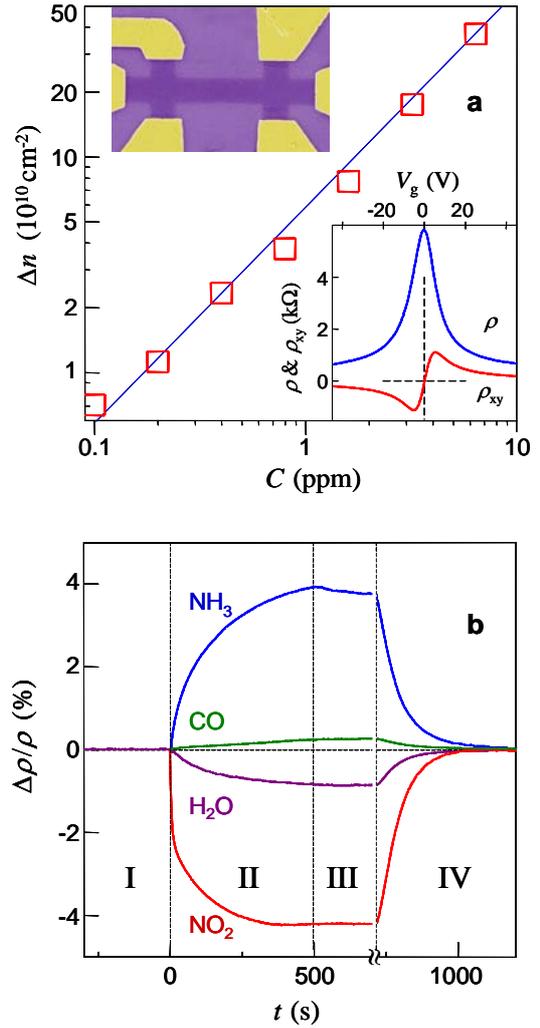

Figure 1. Sensitivity of graphene to chemical doping. (a) – Concentration $\Delta n$ of chemically-induced charge carriers in single-layer graphene exposed to different concentrations $C$ of $NO_2$. Upper inset: scanning-electron micrograph of this device (in false colours matching those seen in visible optics). The scale of the micrograph is given by the width of the Hall bar, which is 1μm. Lower inset: Characterisation of the graphene device by using the electric field effect. By applying positive (negative) $V_g$ between the Si wafer and graphene, we induced electrons (holes) in graphene in concentrations $n = \alpha \cdot V_g$. The coefficient $\alpha \approx 7.2 \cdot 10^{10} cm^{-2}/V$ was found from Hall effect measurements [6-9]. To measure Hall resistivity $\rho_{xy}$, $B$ =1T was applied perpendicular to graphene's surface. (b) – Changes in resistivity $\rho$ at zero $B$ caused by graphene's exposure to various gases diluted in concentration 1 ppm. The positive (negative) sign of changes is chosen here to indicate electron (hole) doping. Region I – the device is in vacuum prior to its exposure; II – exposure to a 5 litre volume of a diluted chemical; III – evacuation of the experimental setup; and IV – annealing at 150ºC. The response time was limited by our gas-handling system and a several-second delay in our lock-in based measurements. Note that the annealing caused an initial spike-like response in $\rho$, which lasted for a few minutes and was generally irreproducible. For clarity, this transient region between III and IV is omitted, as indicated in the figure.

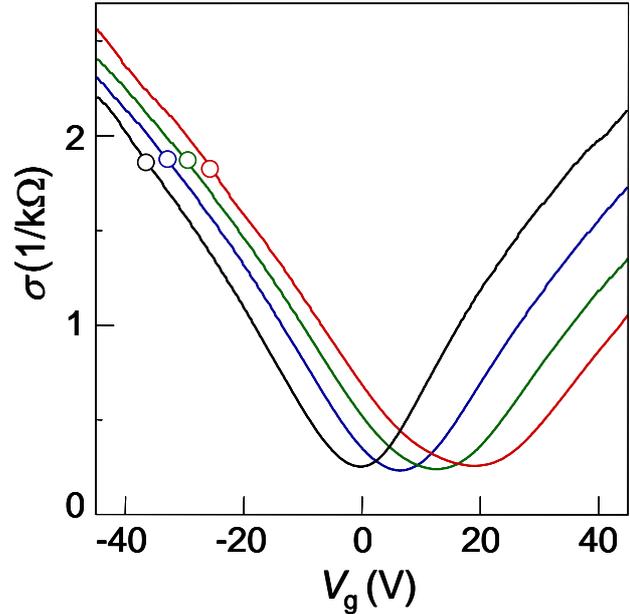

Figure 2. Constant mobility of charge carriers in graphene with increasing chemical doping. Conductivity $\sigma$ of single-layer graphene away from the neutrality point changes approximately linearly with increasing $V_g$ and the steepness of $\sigma(V_g)$-curves (away from the NP) characterizes mobility $\mu$ [6-9]. Doping with NO$_2$ adds holes but also induces charged impurities. The latter apparently do not affect the mobility of either electrons or holes. The parallel shift implies a negligible scattering effect of the charged impurities induced by chemical doping. The open symbols on the curves indicate the same total concentration of holes $n_t$ as found from Hall measurements. The practically constant $\sigma$ for the same $n_t$ yields that the absolute mobility $\mu = \sigma/ne$ as well as the Hall mobility are unaffected by chemical doping. For further analysis and discussions, see Supplementary Information.

The detection limit for solid state gas sensors is usually defined as the minimal concentration that causes a signal exceeding sensors' intrinsic noise [1-4]. In this respect, a typical noise level in our devices $\Delta\rho/\rho \approx 10^{-4}$ (see Fig. 1b) translates into the detection limit of the order of 1 ppb. This already puts graphene on par with other materials used for most sensitive gas sensors [1-4]. Furthermore, to demonstrate the fundamental limit for the sensitivity of graphene-based gas sensors, we optimised our devices and measurements as described in Supplementary Information. In brief, we used high driving currents to suppress the Johnson noise, annealed devices close to NP, where relative changes in $n$ were largest for the same amount of chemical doping, and used few-layer graphene (typically, 3 to 5 layers), which allowed a contact resistance of $\approx 50$ Ohm, much lower than for single-layer graphene. We also employed the Hall geometry that provided the largest response to small changes in $n$ near NP (see lower inset in Fig. 1a). In addition, this measurement geometry minimises the sensitive area to the central region of the Hall cross ($\approx 1\mu$m$^2$ in size) and allows changes in $\rho_{xy}$ to be calibrated directly in terms of charge transfer by comparing the chemically-induced signal with the known response to $V_g$. The latter is important for the low-concentration region where the response of $\rho_{xy}$ to changes in $n$ is steepest but there is no simple relation between $\rho_{xy}$ and $n$.

Figure 3 shows changes in $\rho_{xy}$ caused by adsorption and desorption of individual gas molecules. In these experiments, we first annealed our devices close to the pristine state and then exposed them to a small leak of strongly diluted NO$_2$, which was adjusted so that $\rho_{xy}$ remained nearly constant over several minutes (that is, we tuned the system close to thermal equilibrium where the number of adsorption and desorption events within the Hall cross area was reasonably small). In this regime, the chemically-induced changes in $\rho_{xy}$ were no longer smooth but occurred in a step-like manner as shown in Fig. 3a (blue curve). If we closed the leak and started evacuate the sample space, similar steps occurred but predominantly in the opposite direction (red curve). For finer control of adsorption/desorption rates, we found it useful to slightly adjust temperature while keeping the same leak rate. The characteristic size $\delta R$ of the observed steps in terms of Ohms depended on $B$, the number of graphene layers and, also, varied strongly from one device to another, reflecting the fact that the steepness of $\rho_{xy}$-curves near NP (see Fig. 1a) could be different for different devices [6-9]. However, when the steps were recalibrated in terms of equivalent changes in $V_g$, we found that in order to achieve the typical value of $\delta R$ it always required exactly the same voltage changes $\approx 1.5$mV, for all our 1μm devices and independently of $B$. The latter value corresponds to $\Delta n \approx 10^8 cm^{-2}$ and translates into one electron charge $e$ removed from or added to the area of 1x1μm$^2$ of the Hall cross (note that changes in $\rho_{xy}$ as a function of $V_g$ were smooth, that is, no charge quantization in the devices' transport characteristics occurred – as expected). As a reference, we repeated the same measurements for devices annealed for 2 days at 150C and found no or very few steps (green curve).

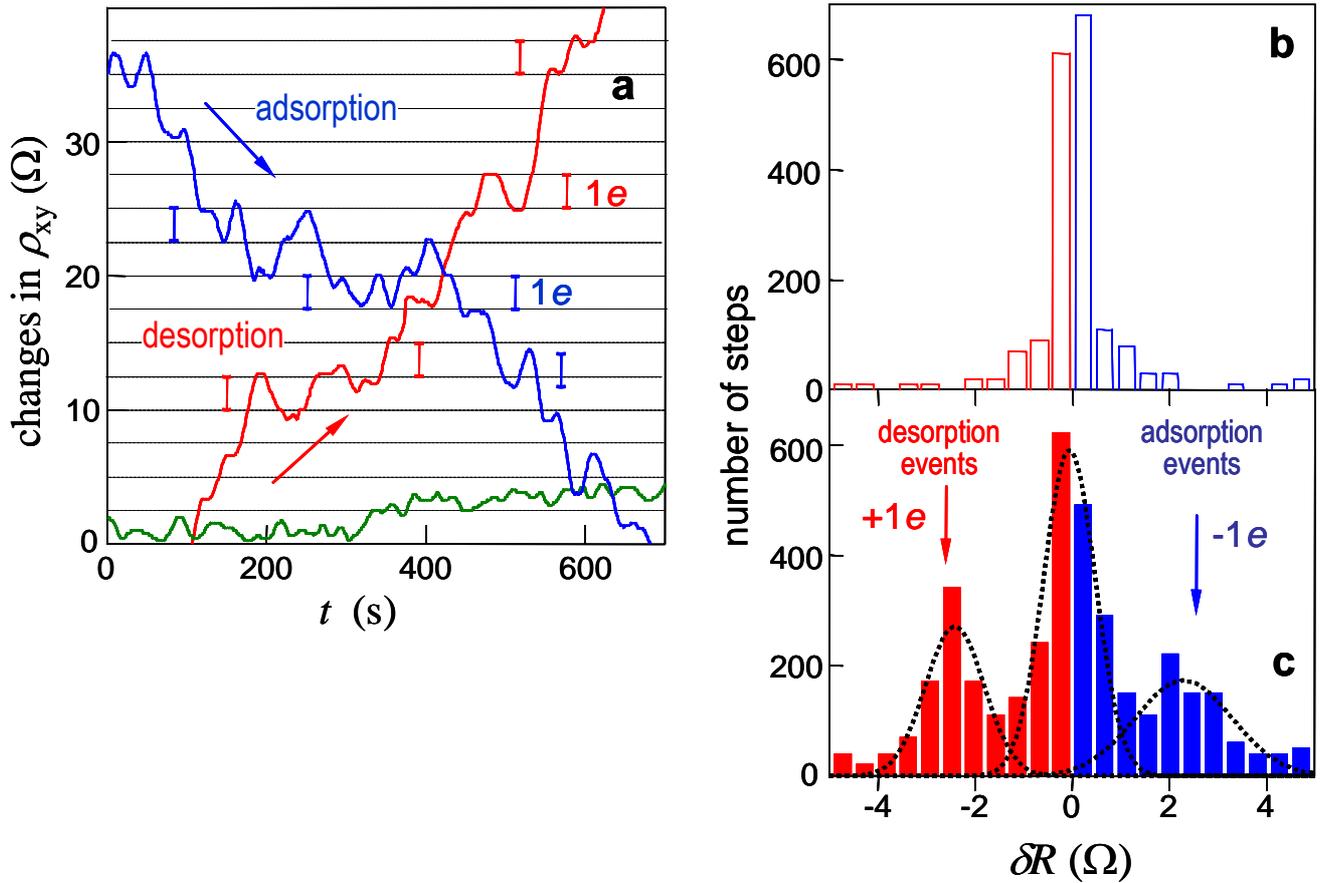

Figure 3. Single-molecule detection. (a) – examples of changes in Hall resistivity observed near the neutrality point ($|n| < 10^{11} cm^{-2}$) during adsorption of strongly diluted $NO_2$ (blue curve) and its desorption in vacuum at 50C (red). The green curve is a reference – the same device thoroughly annealed and then exposed to pure He. The curves are for a 3-layer device in $B = 10T$. The grid lines correspond to changes in $\rho_{xy}$ caused by adding one electron charge $e$ ($\delta R \approx 2.5$ Ohm), as calibrated in independent measurements by varying $V_g$. For the blue curve, the device was exposed to 1 ppm of $NO_2$ leaking at a rate of $\approx 10^{-3}$ mbar·l/s. (b,c) - Statistical distribution of step heights $\delta R$ in this device without its exposure to $NO_2$ (in helium) (b) and during a slow desorption of $NO_2$ (c). For this analysis, all changes in $\rho_{xy}$ larger than 0.5 Ohm and quicker than 10s (lock-in time constant was 1s making the response time of $\approx 6s$) were recorded as individual steps. The dotted curves are automated Gaussian fits (see Supplementary Information).

The curves shown in Fig 3a clearly suggest individual adsorption and desorption events but statistical analysis is required to prove this. To this end, we recorded a large number of curves such as that in Fig. 3a ($\approx 100$ hours on continuous recording). The resulting histograms with and without exposure to $NO_2$ are plotted in Fig. 3b,c (histogram for another device is shown in Supplementary Information). The reference curves exhibited many small (positive and negative) steps, which gave rise to a "noise peak" at small $\delta R$. Large steps were rare. On the contrary, slow adsorption of $NO_2$ or its subsequent desorption led to many large, single-electron steps. The steps were not equal in size, as expected, because gas molecules could be adsorbed anywhere including fringes of the sensitive area, which should result in varying contributions. Moreover, because of a finite time constant (1 sec) used in these sensitive measurements, random resistance fluctuations could overlap with individual steps either enhancing or reducing them and, also, different events could overlap in time occasionally (like the largest step on the red curve in Fig. 3a, which has a quadruple height). The corresponding histogram (Fig. 3c) shows the same "noise peak" as the reference in Fig. 3b but, in addition, there appear two extra maxima that are centred at a value of $\delta R$, which corresponds to removing/adding one acceptor from the detection area. The asymmetry in the statistical distribution in Fig. 3c corresponds to the fact that single-acceptor steps occur predominantly in one direction, that is, $NO_2$ on-average desorbs from graphene's surface in this particular experiment. The observed behaviour leaves no doubt that the changes in graphene conductivity during chemical exposure were quantized, with each event signalling adsorption or desorption of a single $NO_2$ molecule. Similar behaviour was also observed for the case of $NH_3$.

To conclude, graphene-based gas sensors allow the ultimate sensitivity such that the adsorption of individual gas molecules could be detected for the first time. Large arrays of such sensors would increase the catchment area [16], allowing higher sensitivity for short-time exposures and the detection of active (toxic) gases in as minute concentrations as practically desirable. The epitaxial growth of few-layer graphene [17,18] offers a realistic promise of mass production of such devices. Our experiments also show that graphene is sufficiently electronically quiet to be used in single-electron detectors operational at room temperature [19] and in ultra-sensitive sensors of magnetic field or mechanical strain [20], in which the resolution is often limited by $1/f$-noise. Equally important [21,22] is the demonstrated possibility of chemical doping of graphene by both electrons and holes in high concentrations without deterioration of its mobility. This should allow microfabrication of $p$-$n$ junctions, which attract significant interest from the point of view of both fundamental physics and applications. Despite its short history, graphene is considered to be a promising material for electronics by both academic and industrial researchers [6,17,22], and the possibility of its chemical doping improves further the prospects of graphene-based electronics.

We thank Sankar Das Sarma, Allan MacDonald and Vladimir Falko for illuminating discussions. This work was supported by EPSRC (UK) and the Royal Society. M.I.K. acknowledges financial support from FOM (Netherlands). Correspondence and requests for materials should be addressed to K.S.N.

# SUPPLEMENTARY INFORMATION

**Experimental Procedures**

We employed low-frequency (30 to 300 Hz) lock-in measurements and used relatively high driving currents of $\approx 30$ μA/μm. The latter suppressed any voltage noise, so that the remaining fluctuations in the measured resistance were intrinsic, that is, due to thermal switching of unstable defects [5]. Switching defects are known to lead to telegraph noise or, if many such defects are present, to $1/f$-noise, which fundamentally limits the sensitivity of all thin-film sensors at room temperature [5]. In this respect, graphene devices were found to exhibit an exceptionally low level of intrinsic noise, as compared to any other detector based on charge sensitivity (see [19] and references therein). The lowest level of noise was found in devices with the highest mobility (>10,000 cm$^2$/Vs) and the lowest contact resistance. Sensors made from few-layer graphene (3 to 5 layers) were most electrically quiet, probably because their contact resistance could be as low as $\approx 50$ Ohm, as compared with typically $\approx 1$kOhm for our single-layer devices.

To maximize the sensitivity, we tested various regimes and various device's sizes. The maximum signal-to-noise ratio was found for the Hall geometry and measurements at low doping (<$10^{11}$cm$^{-2}$ or $|V_g|$<1V). In this regime, the noise in terms of Ohms was not at its lowest but this was compensated by the steepest response in $\rho_{xy}$ to an induced electric charge (see the lower inset in Fig. 1a). The optimum size was found to be $\approx 1$μm. Smaller devices exhibited higher $1/f$-noise (presumably due to defects at sample edges), whereas larger sizes lead to smaller relative changes in $\rho_{xy}$ in response to the same number of electrons. As an indicator of sufficiently low noise we used the possibility to detect changes with varying gate voltage by less than 1mV. This corresponds to changes of less than one elementary charge $e$ inside the sensitive area of the Hall cross of 1x1μm$^2$ in size.

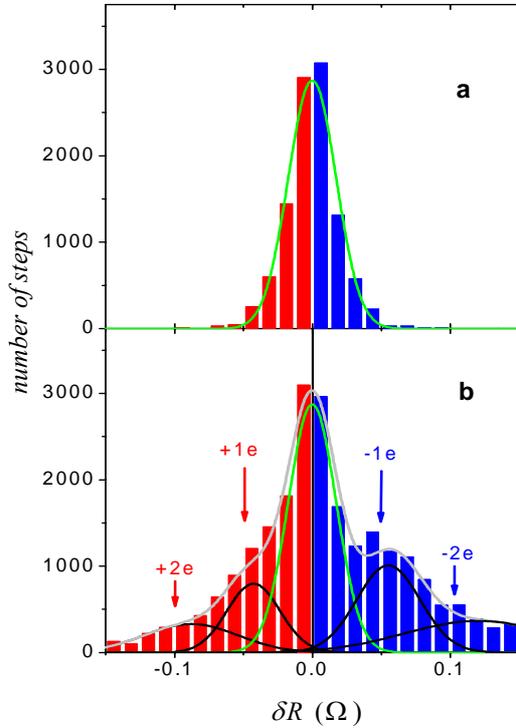

Figure S1. Statistical distribution of step heights $\delta R$ in a (5-7)-layer device during its exposure to pure helium (**a**) and a small leak (10$^{-3}$mbar·l/s) of NO$_2$ diluted in a concentration of 1 ppm (**b**). An example of the raw data is shown by the blue curve in Fig. 3a. Red and blue bars indicate steps in the opposite directions (desorption and adsorption events, respectively). The histogram in (**a**) was first fitted by a Gaussian curve (green). Then, assuming that the noise peak does not change, the remaining statistical distribution was fitted by 4 Gaussian curves (black) allowing all four amplitudes and positions to be chosen automatically by the Origin-7.0 fitting routine. The resulting total of 5 Gaussians accurately fits the whole distribution (grey curve). Three Gaussians also give a reasonable (but less perfect) fit with extra peaks centred at ±0.05Ohm.

**Statistical Distribution of Single-Molecule Steps**

To complement the histograms in Fig. 3 and demonstrate their generality, Fig. S1 shows another example of a histogram for step-like changes in $\rho_{xy}$. These data were obtained for a different device, in a different magnetic field ($B$=4T) and during graphene's exposure to NO$_2$, that is, for the regime of average adsorption, rather than desorption shown in Fig. 3. The 50 times smaller value of the single-electron steps ($\approx 0.05$ Ohm) in this case is due to thicker graphene (5-7 layers), smaller $B$ and a wider transition region near the neutrality point, which leads to less steep changes in $\rho_{xy}$ as a function of $n$. This value of $\approx 0.05$ Ohm was again calibrated using changes in $V_g$ by $\approx 1.4$mV, which adds 1$e$ to the Hall cross area of 1μm$^2$. Due to the weaker response, there is a broad "noise" peak that dominates the statistical distributions in both cases, with and without NO$_2$ exposure. However, it is clear that when the device was exposed to NO$_2$, the statistical distribution became much wider, asymmetric with side wings and cannot be fitted by a single Gaussian. The changes caused by NO$_2$ exposure can only be fitted by adding two additional Gaussian peaks for both negative and positive $\delta R$. However, the automated fitting procedures favour four additional peaks centred at $\approx 0.05$ and $0.1$ Ohm, which exactly corresponds to the transfer of $e$ and $2e$. The $2e$-peak is consistent with events where individual adsorption/desorption steps were not time-resolved and resulted in steps of the double height. The observed asymmetry in the histogram corresponds to the fact that large steps occur predominantly in one direction, that is, the adsorption is stronger than desorption, and graphene's doping gradually increases with time (compare with the asymmetry in Fig. 3c).

**Accumulation of chemical doping**

We found that our graphene devices did not exhibit the saturation in the detected signal during long exposures to small (ppm) concentrations $C$ of active gases. This means that the effect of chemical doping in graphene is cumulative. In the particular experiment shown in Fig. 1b, the apparent saturation observed in region II was found to be caused by a limited amount of gas molecules able to reach the micron-sized sensitive area, because of the competition with other, much larger adsorbing areas in the experimental setup. This is in good agreement with the theory of chemical detectors of a finite size [16]. Figure S2 illustrates the accumulation effect by showing changes in $\rho_{xx}$ and $\rho_{xy}$ as a function of exposure time $t$ for the same sensor as in Fig. 1b but exposed to a constant flow of $NO_2$ and $NH_3$ (in ppm concentrations) rather than to a limited volume of these chemicals as it was the case of Fig. 1b of the main text. In Fig. S2, graphene's doping continues to increase with time $t$ because of the continuous supply of active molecules into the sensitive area (in contrast to the experiment shown in Fig. 1). Within an hour, the device's resistivity changed by 300%. Longer exposures and high $C$ allowed us to reach a doping level up to $\approx 2\times 10^{13} cm^{-2}$. Note that the behaviour in Fig. S2 clearly resembles the corresponding dependences in the lower inset of Fig. 1a but charge carriers in Fig. S2 are induced by chemical rather than electric-field doping. The observed accumulation effect yields that the detection limits for graphene sensors can be exceedingly small during long exposures that allow a sufficient amount of gas molecules to be adsorbed within the sensitive area. Alternatively, large arrays of such sensors would increase the catchment area and should allow a much higher sensitivity also for short-time exposures [16].

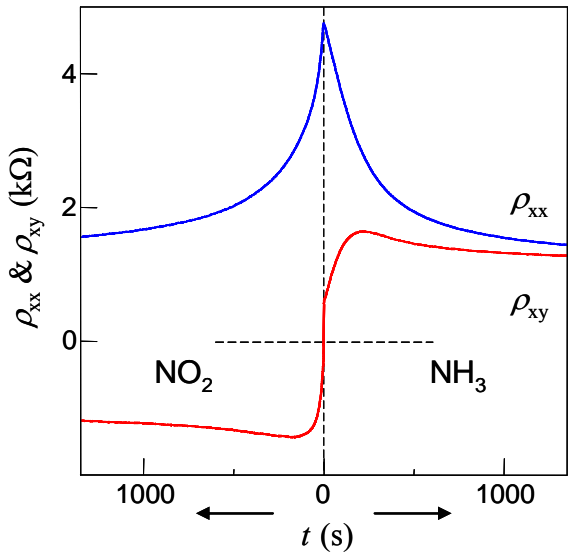

Figure S2. Accumulation of dopants on graphene. Changes in the longitudinal ($\rho_{xx}$) and Hall ($\rho_{xy}$) resistivity of graphene exposed to a continuous supply of strongly-diluted $NH_3$ (right part). After the exposure, the device was annealed close to the pristine state and then exposed to $NO_2$ in exactly the same fashion (left part). Here, measurements of both $\rho_{xx}$ and $\rho_{xy}$ were carried out in field $B=1T$.

The mechanism of chemical doping in graphene is expected to be similar to the one in carbon nanotubes. Unfortunately, the latter remains unexplained and still controversial, being attributed to either charge transfer or changes in scattering rates or changes in contact resistance [3,S1,S2,S3,S4]. Our geometry of four-probe measurements rules out any effect due to electrical contacts, whereas the mobility measurements prove that the charge transfer is the dominant mechanism of chemical sensing. Also, it is believed that the presence of a substrate can be important for chemical sensing in carbon nanotubes. We cannot exclude such influence, although this is rather unlikely for flat graphene, where doping mostly occurs from the top. We also note that hydrocarbon residues on graphene's surface (including remains of electron-beam resist) are practically unavoidable, and we believe that such polymers may effectively "functionalize" graphene, acting as both adsorption sites and intermediaries in charge transfer (see further).

**Constant mobility of charge carriers with increasing chemical doping**

No noticeable changes in $\mu$ with increasing chemical doping were observed in our experiments, as discussed in the main text. In order to estimate quantitatively the extent, to which chemical doping may influence carrier mobility in graphene, we used the following analysis (see Fig. S3). For each level of chemical doping, we measured the dependence of $\sigma$ on $V_g$ (such as in Fig. 2) and the Hall effect in $B =1T$. The latter allowed us to find gate voltages that correspond exactly the same total concentration $n_t =B/e\rho_{xy}$ which combines the concentrations induced by chemical ($N_i$) and electric-field ($n=\alpha V_g$) doping. For example, the symbols in Fig. 2 indicate $n_t \approx 2.7 \times 10^{12}$ cm$^{-2}$. The fact that, for the same $n_t$, $\sigma$ remains unchanged, independently of chemical doping, (Fig. 2) yields that the Hall mobility $\mu =\rho_{xy}/\rho_{xx}B =\sigma/en_t$ does not change. Furthermore, we also calculated the field-effect mobility defined as $\mu =\Delta\sigma/\Delta n$. To this end, the curves were first fitted by linear dependences over an interval of $\pm 10V$. From the found slopes $\Delta\sigma/\Delta V_g$, we extracted the field-effect mobility $\mu =\Delta\sigma/e\alpha\Delta V_g$. An example of the latter for the same $n_t \approx 2.7 \times 10^{12}$ cm$^{-2}$ is plotted as a function of $N_i$ in Fig. S3.

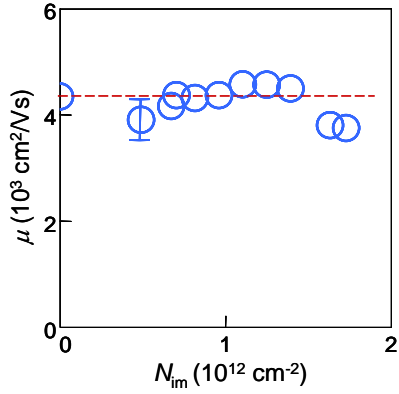

Figure S3. Changes in carrier mobility with increasing the concentration of acceptors induced by $NO_2$ doping

Figs 2 and S3 show that both Hall and field-effect $\mu$ were practically independent of chemical doping. Only for $N_i >> 10^{12}$cm$^{-2}$, we usually found notable changes in the shape of $\sigma(V_g)$-curves, which often became rather deformed. The latter effect remains to be understood, which unfortunately does not allow us to draw quantitative conclusions about the exact behaviour of $\mu$ at very high chemical doping. However, even for $\Delta n \approx 10^{13}$ cm$^{-2}$, we observed the electric-field mobility exceeding 2,000 cm$^2$/Vs, which puts only the lower limit on $\mu$ at such high doping. Also, note a significant broadening of the transition region near NP caused by chemical doping, which is clearly seen on $\sigma(V_g)$-curves in Fig. 2. This broadening could in principle be attributed to an increasingly inhomogeneous distribution of dopants [6,13]. However, such a strong broadening was found to be specific for $NO_2$ and can be explained by two types of acceptor levels (monomers and dimers of $NO_2$) [S5]. This broadening is irrelevant for our main conclusion that graphene's mobility is unaffected by chemical doping, because $\mu$ is defined at high $n$, away from NP [6-9].

Fig. S3 yields that charged impurities in concentration $N_i \approx 10^{12}$ cm$^{-2}$ do not change mobility $\mu \approx 5,000$ cm$^2$/Vs within an experimental accuracy of $\approx 5\%$. This implies that, if all other sources of scattering are eliminated, such a level of chemical doping should still allow $\mu$ as high as $10^5$ cm$^2$/Vs. This value is in strong disagreement (by a factor of 20) with the current theoretical estimates for scattering rates in graphene [11-13], which predict a concentration-independent mobility of $\approx 5,000$ cm$^2$/Vs for charged impurities in concentration $10^{12}$ cm$^{-2}$. Note that these theories take into account the Dirac-like spectrum of graphene, which already results in a strongly reduced scattering in comparison with conventional, Schrödinger-like 2D systems (see below).

There are three possible ways to reconcile the experiment and theory. First, chemical doping can neutralize ionized impurities of the opposite sign, if a mixture of donors and acceptors in a concentration of $\approx 10^{12}$ cm$^{-2}$ is already present at the surface of graphene or in a substrate [S6]. In this case, mobility $\mu$ may even temporarily increase with increasing chemical doping [S6]. However, a large experimental range of $N_i$ over which $\mu$ remains practically unaffected for both electron and hole conductivities (and remains relatively high at $N_i > 10^{13}$ cm$^{-2}$) seems to rule out this mechanism as dominant in our case. Second, absorption sites can be at sample edges or at some distance above a graphene sheet. The former is unlikely for the lack of a sufficient number of broken bonds to accommodate all the dopants along the edges. However, we cannot rule out that a hydrocarbon residue can somehow act as a transfer medium, providing an increased distance between adsorbed impurities and graphene. Indeed, even though our devices were thoroughly cleaned after microfabrication procedures, a thin polymer layer (of about 1nm thick) was observed in AFM and some TEM measurements. This separation is however insufficient [12,13] to explain the observed reduction in scattering rates by a factor of >20. The third possibility is due to absorbed water above or below a graphene sheet, which has a huge dielectric constant $\varepsilon_w =80$ and can provide additional screening [S7]. Indeed, when calculating scattering rates in graphene, it is normally assumed that graphene is neighboured by vacuum and $SiO_2$, a space with an effective dielectric constant $\varepsilon_{eff} = (\varepsilon_{SiO2} + 1)/2 \approx 2.5$ [12,13]. We argue that the presence of a few-nm-thick layer of absorbed water can dramatically increase $\varepsilon_{eff}$ and suppress the scattering contribution of charged impurities below the current detection limit.

It is well known that, unless heated at several hundred C° in high vacuum, all surfaces are covered with absorbed water. For example, $SiO_2$ is normally covered by 2 to 3 nm of water, even in vacuum [S8]. Our analysis of the corresponding electrostatic problem shows that the effective dielectric constant for a graphene sheet that is neighboured by an additional layer of absorbed water with thickness $D$ can be described by $\varepsilon_{eff}(k) \approx [\varepsilon_{SiO2} + 1 + \varepsilon_w \tanh(k_F D)]/2$ where $k_F$ is the Fermi wave vector. For a typical concentration of $10^{12}$ cm$^{-2}$, $\varepsilon_{eff} \approx 10$ and 22 for $D = 1$ and 3nm, respectively. As the scattering rate by charged impurities depends quadratically on $\varepsilon_{eff}$, this additional dielectric screening is sufficient to explain the observed constant mobility with increasing chemical doping. The use of water as a dielectric media suppressing scattering in graphene is an interesting effect that can be used in future to improve the electronic quality of graphene devices.

**On alternative mechanism limiting carrier mobility in graphene**

Our experiments and discussion above show that charged impurities are unlikely to be dominant scatterers in the existing graphene samples. Below we suggest an alternative temperature-independent scattering mechanism but let us first review other possibilities.

It has been shown that scattering on a short-range potential with a radius $R \approx a$ results in low excess resistivity $\rho \approx (h/4e^2) N_i R^2$ where $a$ is the interatomic distance [11-13,S9]. This scattering mechanism can be neglected for any feasible concentration of short-range impurities. Note that, in a normal 2D electron system with a parabolic spectrum, the same concentration of short-range impurities leads to a much higher resistivity $\rho \approx (h/4e^2)(N_i/n)\ln^2(R/\lambda)$ [S10]. One can understand so little scattering on a short-range potential in graphene by using an analogy with the diffraction of light on small obstacles, which becomes inefficient for wavelengths $\lambda \gg R$. This analogy with light is inapplicable for 2D Schrödinger-like electrons because in the latter case a short-range potential always leads to a resonant-like scattering [S9,S10]. On the contrary, for 2D Dirac fermions, the scattering becomes efficient only if an impurity has a bound level at the same energy as that of incident fermions [S9], which would be unusual for graphene because of the Klein tunnelling [6].

To explain the observed values of $\mu$ in graphene and, particularly, its practically constant value with increasing $V_g$ [6-9], a scattering on a long-range Coulomb potential due to charged impurities was invoked [11-13]. Coulomb impurities in a 2D gas of Dirac fermions result in its resistivity $\rho \approx \alpha(h/4e^2)(N_i/n)$ where the coefficient $\alpha$ is predicted to be $\approx 0.2$ [13], which yields $\mu \approx 5,000$ cm$^2$/Vs for $N_i \approx 10^{12}$ cm$^{-2}$. As discussed in the previous section, our experiments prove that chemical doping at $N_i \approx 10^{12}$ cm$^{-2}$ should allow $\mu \approx 10^5$ cm$^2$/Vs, which casts serious doubts that ionized impurities are currently a limiting factor for $\mu$ in graphene.

Therefore, it is sensible to consider alternative scattering mechanisms. To this end, it was experimentally found that graphene is not flat but exhibits random nm-size ripples that involve a large elastic strain of $\approx 1\%$ [14,15]. The influence of such ripples on $\rho$ has not been discussed so far but it was shown that the associated elastic strain effectively results in random vector [6,14] and electric [S11] potentials. The induced vector potential is equivalent to a random sign-changing $B$ exceeding 1 Tesla, which was shown to be sufficient for suppressing weak localization corrections in graphene [6,14]. Below, we show that this random $B$ can induce significant scattering (also, see [S12]).

**Resistivity of a rippled graphene sheet**

Applying the standard procedures for calculating the mean-free time $\tau$ [11-13,S8-S10] but now for the case of a scattering potential with a spinor structure $\vec{V}\vec{\sigma}$, one can write

$$\frac{1}{\tau} \approx \frac{2\pi}{\hbar} N(E_F) \langle \vec{V}_{\vec{q}} \vec{V}_{-\vec{q}} \rangle_{q \approx k_F} \quad (S1)$$

where $N(E_F)$ is the density of states at the Fermi energy and $q$ the wave vector. For a curved surface with the fluctuating height $h(x,y)$ counted from the average plane $z = 0$, the vector potential is proportional to in-plane deformations and, thus, quadratic in the derivatives $\frac{\partial h}{\partial x}, \frac{\partial h}{\partial y}$ (explicit expressions can be found in [S9]; see equations (2)-(5)). This leads to the following expression

$$\langle \vec{V}_{\vec{q}} \vec{V}_{-\vec{q}} \rangle \approx \left(\frac{\hbar v_F}{a}\right)^2 \sum_{\vec{q}_1 \vec{q}_2} \langle h_{\vec{q}-\vec{q}_1} h_{\vec{q}_1} h_{-\vec{q}+\vec{q}_2} h_{-\vec{q}_2} \rangle [(\vec{q}-\vec{q}_1)\cdot\vec{q}_1][(\vec{q}-\vec{q}_2)\cdot\vec{q}_2] \quad (S2)$$

where $v_F$ is the Fermi velocity, $a$ the lattice constant and $h_q$ the Fourier coefficients.

To proceed further, one needs specify the nature of ripples, because the correlation function in the right-hand side of (S2) depends on a distribution of elastic strain. To this end, we first assume that the ripples observed in graphene initially appear as a result of thermal fluctuations [S13] Then, using the standard harmonic approximation, it is straightforward to estimate (S2). Indeed, the average potential energy per bending mode $E_{\bar{q}} = \kappa q^4 \langle |h_q|^2 \rangle / 2$ should be equal to $k_B T / 2$ ($\kappa \approx 1$eV is the bending stiffness of graphene [S11]), which yields

$$\langle |h_{\bar{q}}|^2 \rangle = \frac{k_B T}{\kappa q^4} \qquad (S3)$$

Note that thermal fluctuations with small $q$ are extremely soft, which can lead to a crumpling instability, that is, the amplitude of fluctuations normal to the membrane plane would grow linearly with increasing the membrane size [S13]. However, an anharmonic coupling between bending and stretching modes partially suppresses the growth of such fluctuations at small $q$ [S13]

$$\langle |h_{\bar{q}}|^2 \rangle \approx \frac{1}{q^4}\left(\frac{q}{q_0}\right)^\eta \qquad (S4)$$

where $q_0 \approx \sqrt{b/\kappa} \approx 1/a$ is a typical cut-off vector on interatomic distances, $b$ the 2D bulk modulus, $\eta \approx 0.8$ the bending stiffness exponent [S13]. Changes in the asymptotic behaviour happen for a typical wave vector $q^* = q_0 (k_B T/\kappa)^{1/\eta}$, at which expressions (S3) and (S4) become comparable. At room temperature, this yields $q^* \approx 10^{-2}/a$.

Our crucial assumption is that the thermodynamic distribution of ripples becomes static ("quenched") when a graphene sheet is deposited on a substrate at some quench temperature $T_q$ (300K in our case). Indeed, it is reasonable to suggest that during the deposition process graphene sticks to the substrate and cannot adopt a ripple-free configuration or follow exactly the form prescribed by substrate's own roughness [S14].

For carrier concentrations such that $k_F \geq q^*$ (that is always the case of our measurements of $\mu$), we can use (S3) for the pair correlation function and the Wick theorem for the four-$h$ correlation function in (S2), which allows us to find the ripple resistivity as

$$\rho \approx \frac{h}{4e^2} \frac{(k_B T_q / \kappa a)^2}{n} \Lambda \qquad (S5)$$

where the factor $\Lambda$ is of order of unity for $k_F \cong q^*$ and weakly depends on carrier concentration (as $\ln^2(k_F/q^*)$ for $k_F \gg q^*$). The above equation shows that thermodynamically-induced ripples lead to $\mu$ practically independent on $n$, as observed experimentally. Importantly, (S5) also yields $\mu$ of the same order of magnitude as found in graphene (one can interpret $(k_B T_q / \kappa a)^2 \approx 10^{12}$ cm$^{-2}$ as an effective concentration of ripples).

Finally, we note that if ripples have an origin different from the one discussed above (for example, due to intrinsic roughness of the SiO$_2$ substrate [S14]), then in order to calculate their scattering rates, one would have to know an exact distribution of the associated strain [S15]. Furthermore, it is possible that a structural distribution of ripples is dominated by ripples with a short-range scattering potential [S14] but resistivity is still dominated by a minority of thermodynamically-induced ripples with the long-range potential that is the only efficient source of scattering in graphene.